\documentclass[a4paper,fleqn,usenatbib]{mnras}

\usepackage[T1]{fontenc}
\usepackage{ae,aecompl}
\usepackage{ulem}

\usepackage{graphicx}	
\usepackage{amsmath}	
\usepackage{amssymb}	

\title[Formation of Rocky Exomoons]{Formation of Massive Rocky Exomoons by Giant Impact}

\author[A. C. Barr and M. Bruck Syal]{
Amy C. Barr,$^{1}$\thanks{E-mail: amy@psi.edu (ACB)}
Megan Bruck Syal$^{2}$
\\
$^{1}$Planetary Science Institute, 1700 E. Ft. Lowell, Suite 106, Tucson, AZ 85719 USA\\
$^{2}$Lawrence Livermore National Laboratory, PO Box 808 L-16, Livermore, CA 94550 USA\\}

\date{Accepted 2017 January 9. Received 2017 January 9; in original form 2016 November 17}

\pubyear{2017}

\begin{document}
\label{firstpage}
\pagerange{\pageref{firstpage}--\pageref{lastpage}}
\maketitle

\begin{abstract}
The formation of satellites is thought to be a natural by-product of planet formation in our Solar System, and thus, moons of extrasolar planets (exomoons) may be abundant in extrasolar planetary systems, as well.  Exomoons have yet to be discovered.  However, moons larger than 0.1 Earth masses can be detected and characterized using current transit techniques.  Here, we show that collisions between rocky planets with masses between a quarter to ten Earth masses can create impact-generated debris disks that could accrete into moons.  Collisions between like-sized objects, at oblique impact angles, and velocities near escape speed create disks massive enough to form satellites that are dynamically stable against planetary tides.  Impacts of this type onto a superearth between 2 to 7 Earth masses can launch into orbit enough mass to create a satellite large enough to be detected in \textit{Kepler} transit data.  Impact velocity is a crucial controlling factor on disk mass, which has been overlooked in all prior studies of moon formation via planetary collision.  
\end{abstract}




\section{Introduction}
The \textit{Kepler} spacecraft has discovered thousands of extrasolar planets, many of which may be ``superearths,'' planets with a rock/metal composition, but with masses larger than Earth's mass \citep{Borucki2010,Batalha2014,Borucki2016}.  With the discovery of so many extrasolar planets, the question naturally arises: could these planets have moons?  

A wide variety of techniques have been proposed for exomoon searches (see, e.g., \citealt{HellerAbio2014} and \citealt{exomoon_review} for discussion), however transit techniques hold the most promise for characterization of the moon, allowing a determination of both its mass and radius.  A large exomoon causes the planet/moon system to wobble around the system's center of mass \citep{Szabo2006}, which changes the timing between transits, and the duration of transits \citep{Kipping2009b}.  The magnitude of these effects depend on the satellite-to-planet mass ratio ($M_s/M_{pl}$), and other orbital parameters \citep{Sartoretti1999, Kipping2009b}.  Large moons, and moons that are large relative to their parent planet, are most likely to be detected by the Hunt for Exomoons with \textit{Kepler} (HEK) or any other transit-based survey \citep{Kipping2012-1}.  The simplest and most restrictive detection criterion for the HEK study places the detection threshold at moons greater than $\sim0.1$ Earth masses ($M_E=5.98 \times 10^{27}$ grams) \citep{Kipping2012-1}.  

Of the mechanisms of satellite formation operational in our Solar System, planetary collisions between rocky/icy bodies are the most likely to create systems with large $M_s/M_{pl}$.  The two planet/moon systems in our Solar System with the highest values of this ratio are Earth's Moon ($M_s/M_{pl} \sim 0.012$), and Pluto/Charon ($M_s/M_{pl} \sim 0.12$), which are both thought to have formed from  debris created by a collision of two large objects \citep{HartmannDavis, CameronWard, CanupAsphaug, Canup2004, Canup2005, Canup2011, KenyonBromley2013}.  These types of collisions are thought to be common during the late stages of the formation of terrestrial planets, both in our Solar System, and others \citep{Agnor1999, Morishima2010, Morby2012, Chambers2013}.  Depending on the speed of the collision, the impact angle, and the ratio between the masses of the ``impactor'' (smaller body) and the ``target'' (larger body),  the collision can result in a merger, the formation of an impact-generated disk (from which a moon or moons could form), stripping of the mantle of the target \cite{Asphaug:2006aa}, or a disruptive collision resulting in a cloud of debris (see, e.g., \citealt{Leinhardt2012I, Leinhardt2012II}).  

To date, most numerical simulations of giant impacts have been performed with the goal of reproducing the mass, angular momentum, and composition of the Earth-Moon system (e.g., \citealt{BenzMoon, CanupAsphaug, Canup2004, Elser2011, Canup2012, StewartCuk2012, ReuferMoon2012, Meier2014, Hyodo}) and Pluto-Charon system \citep{Canup2005, Canup2011}.  Very few simulations of satellite formation via giant impact have been performed for impacts involving a total mass, $M_T>M_E$.  Not much is known about how the process scales with $M_T$, or with impact velocity ($v_{imp}$), angle ($\theta$), or impactor-to-system mass ratio ($\gamma$).  Moreover, there is no analytic scaling relationship to predict the outcome of an impact between two planet-scale objects in the parameter space conducive to forming large satellites.  Recent work shows that the creation of large $M_s/M_{pl}$ is not assured in a giant impact: depending on impact geometry and speed,  giant impacts can also yield systems with much smaller $M_{s}/M_{pl}$, such as Phobos and Deimos at Mars \citep{CitronMarsMoons,Rosenblatt2016} and the small ice-rich moons of the Kuiper Belt Object Haumea \citep{LeinhardtHaumea}.   

Here, we report the results of the first hydrocode simulations of the formation of impact-generated debris disks around superearth planets.  Because the accretion of moons in impact-generated disks has only been studied in detail for Earth-mass planets,  we consider the mass of the disk to be the hard upper limit to the mass of moon that could be created in an impact \citep{Elser2011}.  We show that the disk mass depends on impact velocity, which has been overlooked in existing studies, which only consider $v_{imp} \sim 1$ to 1.4 times the mutual escape velocity of the system.  The impact angle and  impactor-to-system mass ratio are also shown to have systematic effects on the disk mass. To date, such relationships have only been partially explored, and only within the context of the impact that formed Earth's Moon.  Finally, we suggest avenues for future work, including additional hydrocode simulations to further constrain how disk mass varies with impact conditions, the development of disk evolution models suitable for the high-temperature, massive disks produced in impacts between superearths.

\section{Background}
\subsection{Disk Mass}
\begin{figure}
\centerline{\includegraphics[width=70mm]{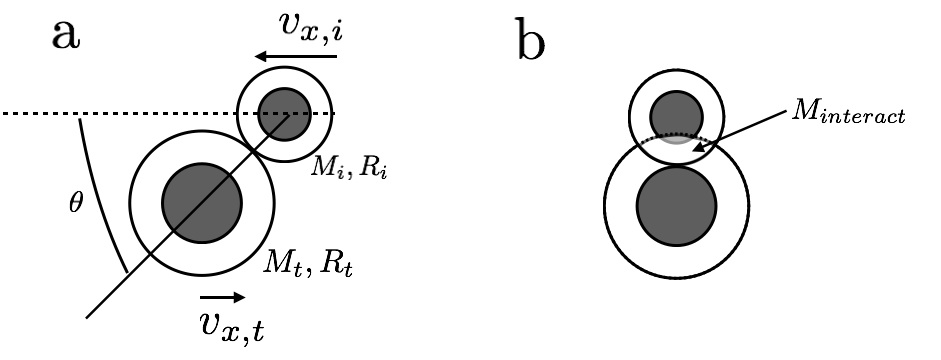}}
\caption{(a) Geometry of a planetary collision, at the point of impact.  A fully differentiated rock (white) and metal (gray) impactor of mass $M_i$ and radius $R_i$ strikes a fully differentiated rock/metal target of mass $M_t$ and radius $R_t$.  The
impact velocity is partitioned between the $\hat{x}$ velocity of the impactor and target to keep the center of mass of the system close to the center of the computational domain.  The diagram also shows the definition of the impact angle, $\theta$.  (b) Definition of $M_{interact}$, the amount of mass in the lens-shaped region of overlap between the impactor and target. \label{fig:geometry}}
\end{figure}

The mass of a satellite formed in a collision depends on the mass of disk material, $M_o$, its angular momentum, $L_o$, and the efficiency with which that material is accreted into one or more final moons \citep{IdaCanupStewart, Canup2008}.  The mass and angular momentum of material launched into orbit depend on the impact geometry, velocity, and objects' compositions.  Figure \ref{fig:geometry}a illustrates the geometry of the initial condition of a planet-planet collision, and the definition of the impact angle, $\theta$.  Impacts that create massive disks, conducive to forming massive satellites, usually occur at close to the mutual escape velocity of the system, $v_{esc,sys}=\sqrt{2G(M_t+M_i)/(R_t+R_i)}$, where $M$ and $R$ are the mass and radius of the target and impactor.  The quantity $\gamma = M_i/(M_i+M_t)$, expresses the ratio between the mass of the impactor to the total mass involved in the collision.

Previous numerical simulations have constrained the mass of impact-generated disk created for $M_T \sim M_E$, $0.1 \lesssim \gamma \lesssim 0.2$, moderate impact angles, and $1 \lesssim v_{imp}/v_{esc} \lesssim 1.4$.  Their results show that the disk mass, $M_o$, depends on the mass of the lens-shaped region representing the overlap between the target and impactor, $M_{interact}$ \citep{Canup2008, Leinhardt2012I}, 
\begin{equation}
\frac{M_o}{M_T} \sim C_{\gamma} \bigg(\frac{M_i - M_{interact}}{M_T}\bigg)^2. \label{eq:diskmass}
\end{equation}
The factor $C_{\gamma} \sim 2.8 (0.1/\gamma)^{1.25}$ is determined empirically based on the results of impact simulations \citep{Canup2008}.  If equation (\ref{eq:diskmass}) holds for a wide range of impact velocities and angles, it would be possible to predict $M_o,$ and thus the maximum $M_s$, without having to simulate the impact.  Unfortunately, the value of $C_{\gamma}$ has only been determined for a narrow range of impact conditions.  Its applicability to impacts beyond these ranges is unclear.  


\subsection{Moon Mass}
After the collision, the debris cloud quickly collapses to a disk \citep{WardCameron1978}, which extends from the surface of the planet, to well beyond the Roche limit, the distance from the planet where tidal forces become too weak to prevent accretion.  The Roche limit lies at $a_R =[3\rho_{pl}/\rho_{s}]^{1/3}R_{pl}$, where $R_{pl}$ is the planet radius, $\rho_{pl}$ is the density of the planet, and $\rho_s$ is the density of disk material \citep{MurrayDermott, Elser2011}.  For the Earth, $a_R \sim 2.9 R_E$, where the radius of the Earth $R_E=6371$ km.  For the disk temperatures and densities associated with the Moon-forming impact, the portion of the disk inside the Roche limit exists as a two-phase liquid/vapor mixture \citep{ThompsonStevenson, Ward2011, Charnoz2015}.  The disk spreads inward, toward the planet, and outward beyond the Roche limit on a time scale that depends on the balance between heating and cooling processes in the disk \citep{Ward2011, Charnoz2015}.  

At present, it is only possible to provide an {\it upper limit} on the mass of moons that could be created from a planetary collision.  Collisions between bodies with masses much greater than $M_E$ involve much higher impact velocities and create disks with higher temperatures and more vapor than those appropriate for the formation of Earth's Moon.  Although the thermal and physical evolution of an impact-generated silicate melt/vapor disk has been studied in detail for Moon-forming conditions, very little is known about how much more massive or hotter disks might behave.

Because of the complexity of disk models and disk processes, most studies of the sweep-up of debris post-impact assume that the disk material has simply condensed into solid particles (e.g., \citealt{IdaCanupStewart, Kokubo2000, Hyodo}); accretion is then governed solely by gravitational forces.  In a purely particulate disks with $M_o/M_{pl} > 0.03$, of the type created in the Moon-forming impact, accretion of a single satellite is favored \citep{IdaCanupStewart, Kokubo2000, Hyodo}, and its mass ($M_L$) is proportional to the ratio of disk angular momentum to the angular momentum of an orbit at the Roche limit \citep{IdaCanupStewart,Kokubo2000},
\begin{equation}
\frac{M_L}{M_o} \approx a \bigg(\frac{L_o}{M_o \sqrt{GM_{pl} a_R}}\bigg) - b - c \bigg(\frac{M_{esc}}{M_o}\bigg),
\end{equation}
where $a=1.9$, $b=1.15$, and $c=1.9$ are numerical coefficients obtained from fitting the outcomes of many $N$-body simulations, $L_o$ is the angular momentum of orbiting material, and the amount of mass that escapes the system during accretion, $M_{esc} \leq 5$\%.  For disks with $0.003 < M_o/M_{pl} < 0.03$, a handful of $N$-body simulations for a narrow range of $L_o$ show that formation of two satellites is favored \citep{Hyodo}, especially in disks with low angular momentum.  The mass of the first satellite is proportional to $M_o^2$.  The two satellites can be comparable in mass, or quite dissimilar in size, and can form in a 2:1 mean motion resonance, or be co-orbital \citep{Kokubo2000, Hyodo}.  For $M_o/M_{pl} < 0.003$, $N$-body simulations of disks with low mass but constant surface-mass density (not a realistic assumption for impact-generated disks) show that $M_L/M_{pl} \sim 2200(M_o/M_{pl})^3$ \citep{CridaCharnoz}. 

At present, only two studies have been able to simulate the coupled evolution of the liquid/vapor disk and gravitational sweep-up of debris outside the Roche limit \citep{SalmonCanup, SalmonCanup2014}.  For disks created by the oblique, ``canonical'' Moon-forming impact identified by \citet{Canup2004}, their study finds $a=1.14$, $b=0.67$, and $c=2.3$, thus predicting slightly lower moon masses than prior studies.  These coefficients also hold for relatively higher-temperature and more massive disks created by the head-on collision of two $\sim 0.5 M_E$ objects \citep{SalmonCanup2014}, but their applicability to superearth collisions has not been established.

\section{Methods}
\subsection{Hydrocode Simulations}
We simulate impacts using the Eulerian/Adaptive Mesh Refinement CTH shock physics code \citep{McGlaun}, which solves the governing equations of shock wave propagation and deformation.  CTH has been recently modified to include self-gravity \citep{Crawford1999, Crawford2006}.  CTH has been widely used by the planetary science community to simulate local- and planetary-scale impacts (e.g., \citealt{Pierazzo1997, Barr:2011aa, MeganNGeo}), including the Moon-forming impact \citep{m-moon}.  

We use an improved version of the semi-analytic equation of state, ANEOS \citep{ThompsonLauson1972}, which includes a description of molecular (as opposed to purely atomic) vapor \citep{MeloshMANEOS} with coefficients for iron and rock (dunite composed of 100\% forsterite, Mg$_2$SiO$_4$) from \citet{ThompsonLauson1972}, \citet{CanupAsphaug}, and \citet{m-moon}.  The impacts are simulated in a fully three-dimensional Cartesian domain, $\sim125R_E$ on a side to ensure that no material leaves the domain.  We use adaptive meshing to concentrate numerical resolution in locations of high density (e.g., \citealt{CrawfordKipp}).  Our scheme puts roughly equal mass in each grid cell, mimicking the way that smoothed particle hydrodynamics (SPH) divides mass into particles \citep{m-moon}.  The smallest element is $\sim 0.01R_t$ to $0.05R_t$, or a few tens to hundreds of kilometers.  At the endpoint of a simulation, the planet, disk, and any orbiting clumps of material are typically represented by $10^6$ to $10^7$ elements.  

\subsection{Initial Conditions}
The bodies are initially $\sim(R_t+R_i)$ apart.  Each body moves toward the other in the $\hat{x}$ direction at a velocity chosen so that the center of mass remains close to the center of the domain.  In this initial exploratory study, we do not include the effect of pre-impact rotation, which has been shown to affect $M_o$ \citep{Canup2008}.  Different impact angles are obtained by changing the initial location of the impactor in the $\hat{y}$ direction.  

The target and the impactor are 30\% iron and 70\% dunite by mass \citep{Canup2004, MarcusRocky2009}.  We specify an initial estimate for the central temperature and surface temperature for both the impactor and the target.  In its first time step, CTH uses a built-in subroutine to determine initial pressure and density profiles consistent with the equations of state, and the estimated central and surface temperatures.  Initial masses and radii for the target and impactor are used to calculate $M_{interact}$, the amount of impactor and target material contained in the lens defined by Figure \ref{fig:geometry}b. We also calculate the proportions of rock and metal contained in the overlapping lens.  For the simulation depicted in Figure \ref{fig:angle5}, $M_{interact}/M_T \sim 10^{-3}$.  

\subsection{Analysis}
\begin{figure}
\centerline{\includegraphics[width=80mm]{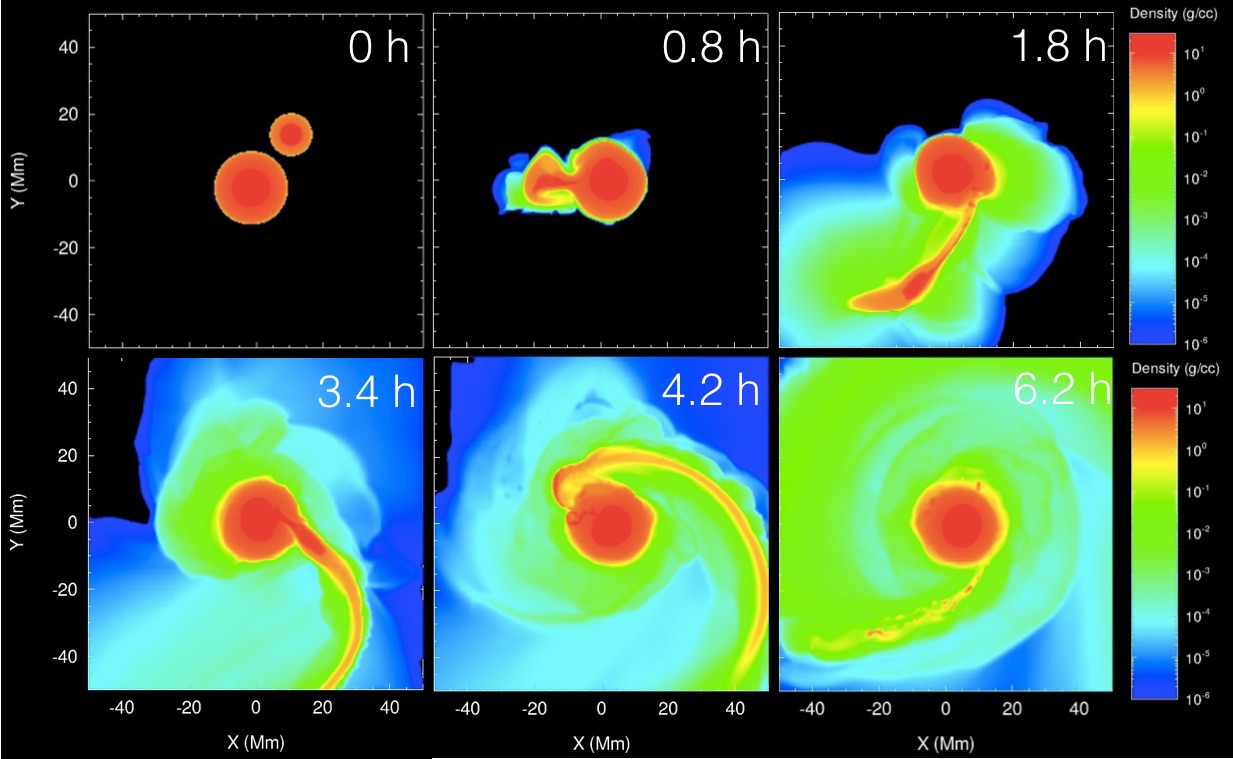}}
\caption{CTH simulation of a $M_t=6.38M_E$ planet colliding with a $M_i=0.82M_E$ impactor at $v_{imp}/v_{esc}= 0.7$, at $70^{\circ}$.  Both objects are composed of 70\% dunite and 30\% iron by mass.  Colours indicate densities on a logarithmic scale, with orange/red indicating $\rho > 1 $ g/cm$^3$ (solid material), blue/green indicating $\rho \sim 10^{-3}$ g/cm$^3$ (vapor), and dark blue indicating $\rho \sim 10^{-6}$ g/cm$^3$ (diffuse vapor).  The collision yields an orbiting disk of mass $M_o\sim 0.15 M_E$.  If all of the disk material accumulates into a single moon, the moon has a radius $R_L \sim 3570$ km (0.58 $R_E$).  The final mass of the planet is $7 M_E$. \label{fig:angle5}}
\end{figure}

Figure \ref{fig:angle5} illustrates the first six hours of an example CTH impact simulation between two rock/metal planets.  The total mass in the collision, $M_T=7.2 M_E$, $\gamma=0.1$, $v_{imp} \sim 0.7 v_{esc}$, and impact angle of 70$^{\circ}$.  The target planet has mass $M_t=6.38M_E$, and it collides with a smaller object $M_i=0.82M_E$.  Early in the simulation, material beneath the impact point is launched into orbit around the planet.  The iron cores of the two objects merge.  These behaviors are similar to those observed in the canonical Moon-forming impact.  The disk of material generated in this collision is denser, with vapor densities $\sim 10^{-2}$ g/cm$^3$, compared to $\sim 10^{-4}$ g/cm$^3$ at comparable times in the Moon-forming impact (see, e.g., Figure 4 of \citet{m-moon} for comparison).

\begin{figure}
\centerline{\includegraphics[width=80mm]{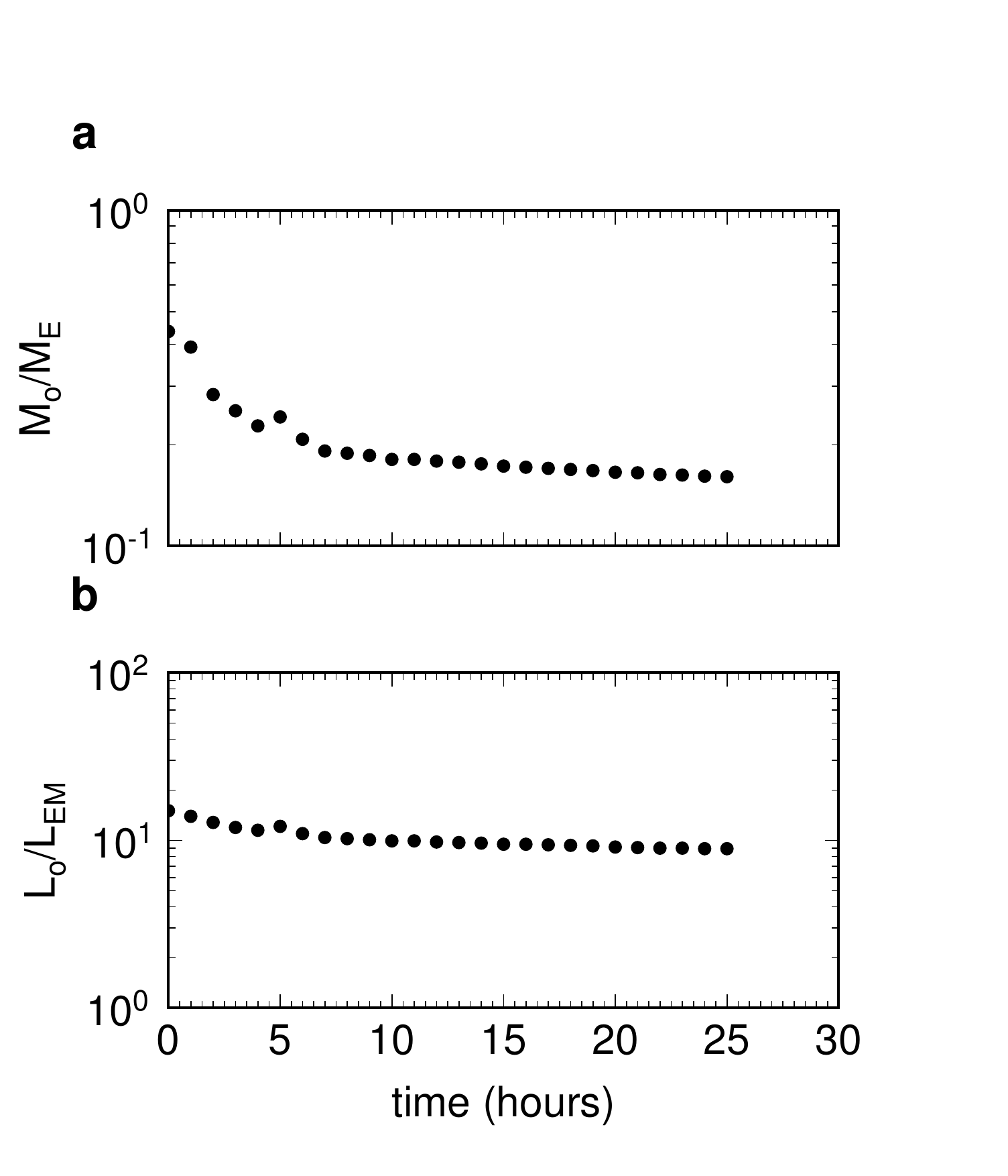}}
\caption{(a) Disk mass $M_o/M_E$, and (b) angular momentum of the disk, $L_o$, scaled by the angular momentum of the Earth/Moon system ($L_{EM}=3.5\times10^{41}$ g cm$^2$/s) a function of time for the simulation depicted in Figure \ref{fig:angle5}.  A few hours after the impact, any clumps of material have re-impacted the planet, and the disk spreads, due in part to the numerical viscosity in CTH \citep{m-moon}, onto the planet and outward, resulting in a decrease in $M_o$ for $t>10$ hours.   \label{fig:angle5-timelogs}}
\end{figure}

For each simulation, we track the disk mass ($M_o$) and angular momentum ($L_o$) as a function of time.  Figure \ref{fig:angle5-timelogs} illustrates the evolution of these quantities as a function of time, for the simulation depicted in Figure \ref{fig:angle5}.  The end result of a giant impact simulation is a cloud of material orbiting a central concentration of mass, and it is not immediately clear how to determine which material is part of the ``planet'' and which is part of the ``disk.''  We follow the procedure described by \citet{Canup2001}.  We guess the physical size of the planet, $R_{pl}$, which is usually larger than $R_t$, because some of the impactor material will have merged with the planet.  For parcels outside the planet, we calculate the radius of the equivalent circular orbit, $r_{circ}$.  If $r_{circ} < R_{pl}$, its mass is added to the planet.  This procedure is repeated iteratively until the mass and radius of the planet converge \citep{Canup2001}.  We also determine the mass of material that is gravitationally unbound from the system, $M_{esc}$.  A simulation is considered ``complete'' when $M_o$ and $L_o$ do not change significantly in between time steps. 

The satellite will only survive its earliest tidal evolution if the semi-major axis of the synchronous orbit is greater than the Roche limit \citep{Elser2011}.  The Roche limit is the closest that a strengthless satellite can get to the planet without being pulled apart by the planet's gravity.  Immediately after its accretion, the satellite and the planet will begin to exert gravitational torques on each other.  If the satellite is originally located far from the planet, so that its orbital mean motion ($n$) is greater than the spin frequency of the planet ($\Omega$), the satellite will migrate inward, toward the planet, until $\Omega=n$.  If the satellite is located close to the planet, it will migrate away from the planet until $\Omega=n$ (see e.g., \citet{MurrayDermott} for discussion).  The synchronous rotation point, defined as the location where $\Omega=n$ is,
$r_{sync} = [(GM_{pl})/(\Omega^2)]^{1/3}.$
The spin frequency of the planet can be calculated from the angular momentum of the material judged to be part of the planet: $L_{pl}=I_{pl}\Omega$, where the planet's moment of inertia, $I=\alpha_{pl}M_{pl}R_{pl}^2,$ and $R_{pl}$ is the radius of the final planet.  The quantity $\alpha_{pl}$ is the polar moment of inertia coefficient, which we calculate based on the planet's interior state.  

\section{Results}
\begin{figure}
\centerline{\includegraphics[width=90mm]{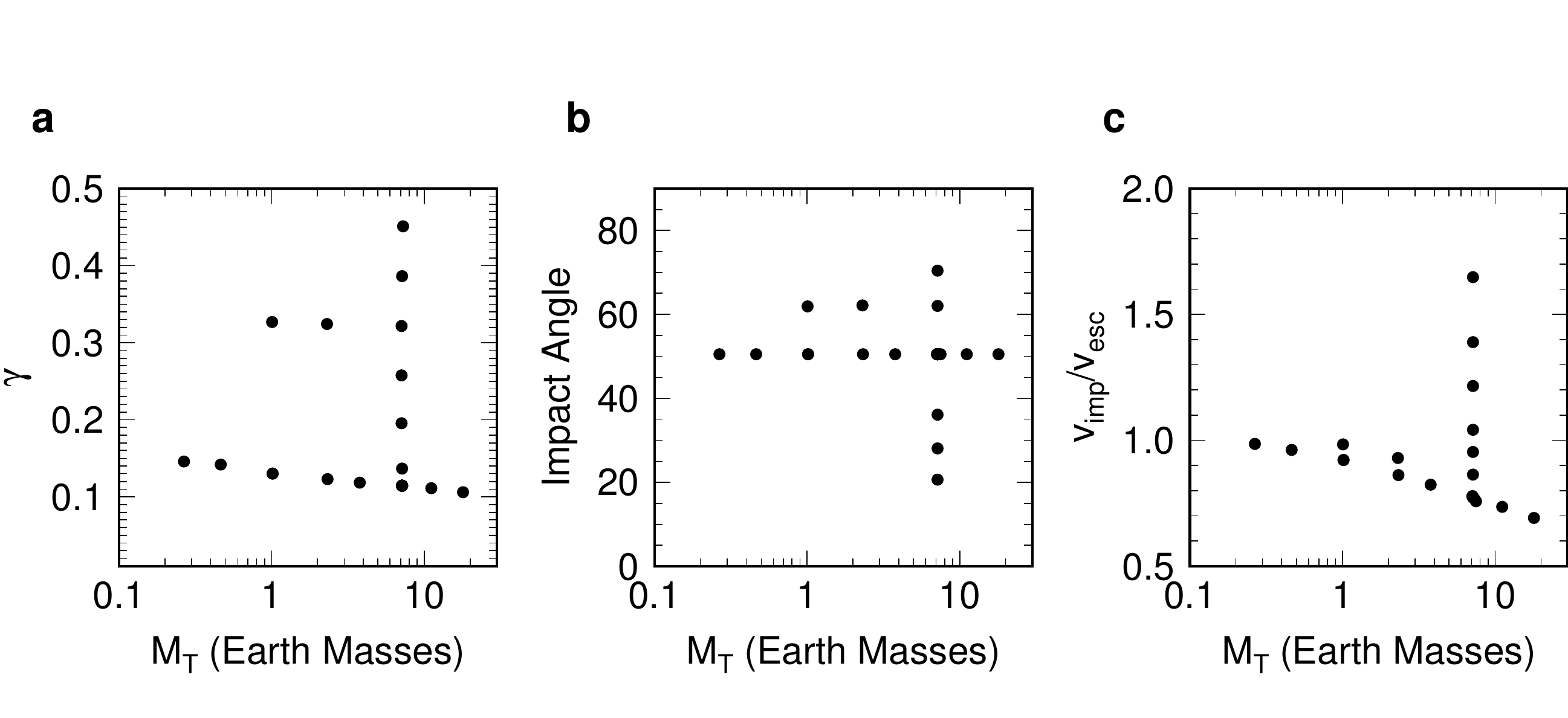}}
\caption{(a) Values of impactor-to-total mass ratios ($\gamma$), (b) impact angles, and (c) velocities explored in this study (points).  \label{fig:paramspace}}
\end{figure}

Figure \ref{fig:paramspace} illustrates the locations of our simulations in $M_T$, impact velocity, $\theta$, and $\gamma$ parameter space.  To begin to understand the conditions that lead to large satellites, we have explored a few ``slices'' through parameter space.  The first slice, depicted as the horizontal rows of points in each panel of Figure \ref{fig:paramspace}, has a geometry identical to the Moon-forming impact, but different $M_T$, ranging from 0.266 to 18.1 Earth masses.  For $M_T=7M_E$, we explored different values of $\gamma$ (ratio of impactor to target mass) while keeping other parameters equal to that for the Moon-forming impact (vertical series of points in Figure \ref{fig:paramspace}a).  Simulations also spanned a range of impact angles, shown in Figure \ref{fig:paramspace}b, and velocities, shown in Figure \ref{fig:paramspace}c. For a full list of the simulations performed for this paper, see Table \ref{tab:sims}.

\subsection{Impact Velocity}
Figure \ref{fig:results}a illustrates how $M_o/M_{pl}$ depends on impact velocity when all other parameters are constrained to the values commonly used for the Moon-forming impact at Earth ($M_T=7M_E, \gamma=0.11$ and impact angle of 50.55$^{\circ}$).  Impacts with $v_{imp}/v_{esc,sys}\sim1$ yield the largest $M_o$ for a given planet mass.  For impact velocities less than the escape velocity, increasing $v_{imp}/v_{esc,sys}$ leads to increasing $M_o$, $L_o$, and larger predicted masses: the disk mass increases by about an order of magnitude between $v_{imp}/v_{esc,sys}$ =  0.774 and 1.042. This alone suggests that Equation (\ref{eq:diskmass}) does not capture all of the relevant physical effects in planetary collisions conducive to forming moons. While these lower-speed impacts are somewhat less probable, requiring special orbital conditions to achieve, we include them here to demonstrate how deviating in either direction from the lunar-forming impact affects results.

When impact velocities exceed $v_{esc,sys}$, the disk mass drops precipitously, by approximately two orders of magnitude between $v_{imp}/v_{esc,sys}$ =  1.042 and 1.647. However, the three values of disk mass we calculated for larger velocities are relatively constant as a function of $v_{imp}$.  This hints at a possible rapid decrease in retained disk mass as impact velocities increase above $v_{esc,sys}$ and then a convergence to similar values as speeds increase further.  It may be possible that two regimes exist: for velocities less than the escape velocity, disk masses increase as $v_{imp}$ increases; for $v_{imp} > v_{esc,sys}$, disk masses decrease and then remain relatively constant as $v_{imp}$ increases. 

\subsection{Impact Angle}
Figure \ref{fig:results}b shows how $M_o/M_{pl}$ varies as a function of impact angle, $\theta$.  Disk mass is highest for the most glancing impacts and decreases as the impacts approach normal incidence. This is consistent with previous work, which found that impact geometry is one of the most important controls on the final disk mass \citep{Canup2004}. As impacts become more oblique, ejecta momentum is directed downrange, away from the center of the target, allowing more material to be injected into orbit. Additionally, peak shock pressures are lower, which can aid in keeping shocked material at lower temperatures and less likely to escape the system. 

Interestingly, for the most head-on case (20$^{\circ}$ from normal incidence), this trend reverses, as disk mass increases. This may be due to a tradeoff between the increased shock pressures and decreased focusing of downrange ejecta: as impacts approach normal incidence (which is statistically unlikely), increases in shock heating and vaporization drive a larger mass of material into orbit. This effect was not reported in previous studies, which have looked at only a limited range of impact angles.

\subsection{Impactor-to-System Mass Ratio}
\begin{figure}
\centerline{\includegraphics[width=80mm]{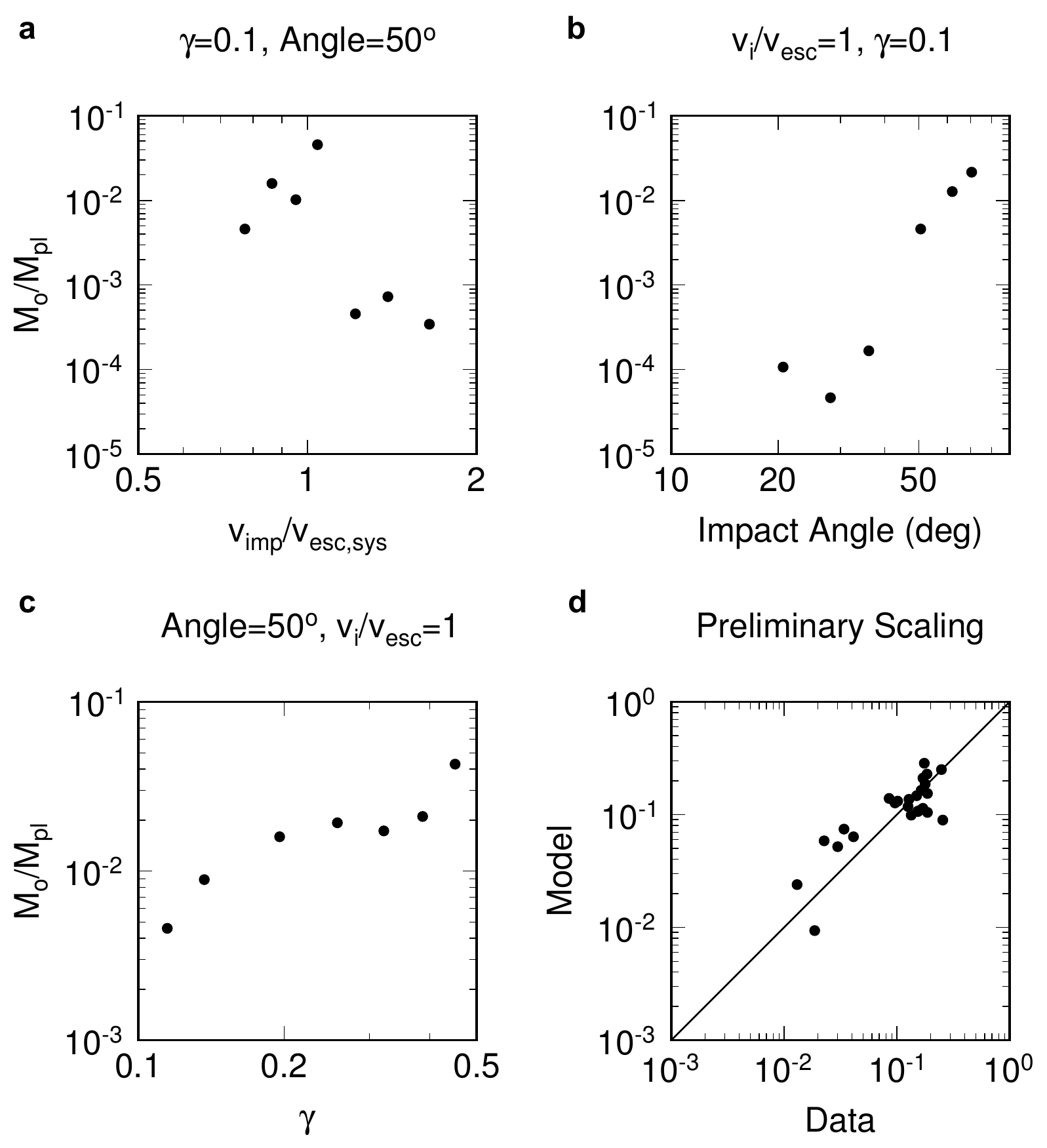}}
\caption{(a) Disk mass divided by final planet mass ($M_o/M_{pl}$) for impacts with varying impact velocity, for $M_T=7M_E$, $\gamma=0.1$, $\theta=50^{\circ}$. (b) Variation in disk mass as a function of impact angle, for $v_{imp}=v_{esc,sys}$ and $\gamma=0.1$. (c) Variation in $M_o/M_{pl}$ as a function of $\gamma$, for $\theta=50^{\circ}$, $v_{imp}=v_{esc,sys}$.  (d) Data fit to the preliminary, suggested scaling relationship.  More simulations are required to decrease the errors on fitting coefficients.
\label{fig:results}}
\end{figure}

Varying the impactor-to-system mass ratio, $\gamma$, also affects $M_o/M_{pl}$ as shown in Figure \ref{fig:results}c. As the impactor and target approach equal size, the disk mass increases: simulations result in an order of magnitude larger $M_o/M_{pl}$ as $\gamma$ increases from 0.115 to 0.45. These effects have not been systematically explored in previous work on satellite-forming impacts, which have generally been confined to either to $\gamma < 0.15$ \citep{Canup2004,StewartCuk2012}, or $\gamma > 0.4$ \citep{Canup2012}. The heightened efficiency of these larger impacts in creating much more massive debris disks have important implications for the upper limits for collisionally formed moons in exoplanetary systems.

\section{Discussion}
Our results are the first to demonstrate the wide diversity of moons expected to form in the more varied set of impact conditions possible within exoplanetary systems. Most importantly, we have shown that it is possible to form exomoons with masses above the theoretical detection limits of the ongoing HEK survey ($M_s>0.1M_E$). Using the mass of the disk as a hard upper limit on the mass of moon that could form \citep{Elser2011}, we show that moons $M_s>0.1M_E$ can form around superearths in collisions with a large impactor-to-system mass ratio, $\gamma$, oblique impact angles, and impact velocities near the system escape speed, $v_{imp}/v_{esc,sys}\sim1$. In our initial suite of impact simulations, we find ten different cases resulting in disk masses $M_s>0.1M_E$, including an oblique, escape-velocity impact between 2.3$M_E$ planet and a $0.75M_E$ impactor, and impacts involving $M_T\sim 7 M_E$ for a variety of impactor sizes, angles, and velocities (Table \ref{tab:sims}).

We find that that disk mass is very sensitive to the impact velocity. The disk mass is maximized for $v_{imp} \sim v_{esc,sys}$, close to the characteristic velocities expected for accretional impacts during the late stages of planet formation \citep{Morishima2010,Chambers2013}.  For $v_{imp} > v_{esc}$, the disk masses fall off as a strong function of $v_{imp}$. The existing scaling relationship for disk mass (Equation \ref{eq:diskmass}) does not consider changes in impact velocity. Additionally, the effects of changing $\gamma$ have only been considered over relatively limited ranges for this value. This motivates the development of a new scaling relationship, which we suggest can take the following form:
 \begin{equation}
\frac{M_o}{M_T} \sim A \gamma^{a} \bigg(\frac{M_i - M_{interact}}{M_T}\bigg)^b \bigg(\frac{v_{imp}}{v_{esc}}\bigg)^c . \label{eq:newscaling}
\end{equation}
The Levenberg-Marquardt method is used to perform a non-linear least squares inversion on the results \citep{NR} to constrain the values of fitting parameters $A$, $a$, $b$, and $c$. Figure \ref{fig:results}d illustrates a comparison between the $M_o/M_T$ data and Equation \ref{eq:newscaling} with $A\sim 0.28$, $a\sim -2.7$, $b\sim 3.7$, and $c\sim-2.7$. The 2-$\sigma$ error bars on the fitting parameters are large, about 50\% the magnitude of the coefficients themselves, thus, it is clear that more simulations will be required to obtain better estimates of the fitting coefficients. Alternatively, a different scaling relationship, perhaps with multiple regimes, may be necessary to account for the full array of conditions encountered in exoplanetary giant impacts. 

\begin{figure}
\centerline{\includegraphics[width=90mm]{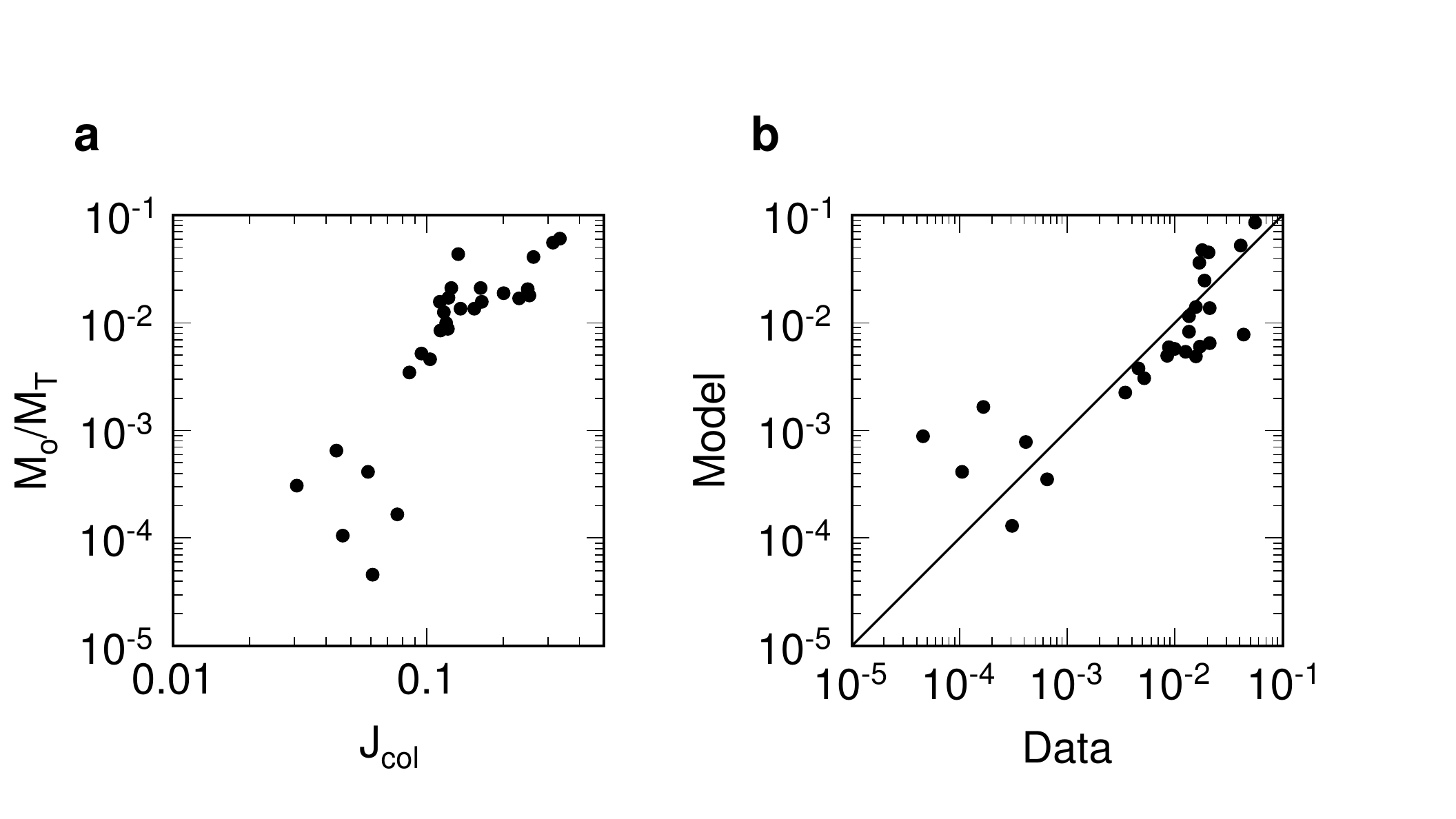}}
\caption{(a) Variation in disk-to-total mass ratio, $M_o/M_T$, as a function of the normalized angular momentum of the collision (see equation \ref{eq:Jcol}), which incorporates $v_{imp}/v_{esc}$, $\theta$, and $\gamma$. (b) Comparison between data and a fit to the data, $M_o/M_T=2.15 J_{col}^{2.79}$.
\label{fig:Jcol}}
\end{figure}

An alternative means of predicting $M_o$ would be to look at how the ratio between the disk mass and the {\it total} mass, $M_o/M_T$, varies with the angular momentum of the collision.  The results of candidate Pluto/Charon impacts from \citet{Canup2005} show that $M_o/M_T$ is proportional to $J_{col}$, the normalized angular momentum of the collision,
\begin{equation}
J_{col}=\sqrt{2} f(\gamma) \sin{\theta} \bigg(\frac{v_{imp}}{v_{esc}}\bigg), \label{eq:Jcol}
\end{equation}
where $f(\gamma)=\gamma(1-\gamma) [\gamma^{1/3} + (1-\gamma)^{1/3}]^{1/2}$.  Figure \ref{fig:Jcol}a illustrates how $M_o/M_T$ varies as a function of $J_{col}$.  A fit to our data (illustrated in Figure \ref{fig:Jcol}b) shows that $M_o/M_T \approx 2.15 J_{col}^{2.79}$ over several orders of magnitude in $M_o/M_T$.  However, it is clear that further simulations will be needed to determine how the disk mass varies with collision geometry.  Additional simulations we are presently undertaking will allow us to better-constrain these relationships.

\begin{figure}
\centerline{\includegraphics[width=80mm]{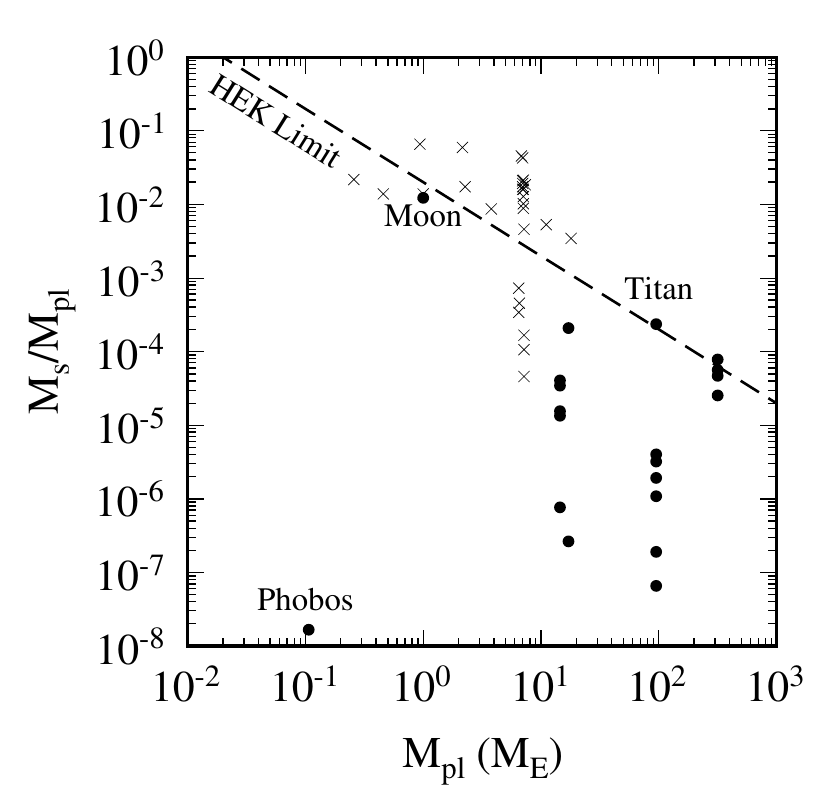}}
\caption{Satellites in our Solar System (dots) plotted alongside simulation results from this study (x symbols), as a function of host planet mass, $M_{pl}$ (in Earth masses, $M_E$), and final satellite-to-planet mass ratio, $M_s/M_{pl}$. The approximate detection limits for exomoons by the \textit{Kepler} spacecraft are indicated by the ``HEK Limit'' line. Note that many of the impact-generated moons modeled in this work exceed the HEK survey detectability threshold.
\label{fig:moons}}
\end{figure}

As demonstrated in Figure \ref{fig:moons}, we have already created a suite of objects that could be detectable with HEK.  The largest satellite-to-planet mass ratio we have obtained so far is 6\%, created in an impact with $M_T\sim M_E$, $\gamma=0.33$, and $\theta=62^{\circ}$ at $v_{imp}/v_{esc} \sim 1$, not dissimilar from that inferred for the formation of the Pluto/Charon system \citep{Canup2005,Canup2011}. The satellite orbits an Earth-mass planet, but it is five times the mass of the Moon.  This satellite could be observable in the HEK survey, but it is only marginally stable against planetary tides, with $r_{sync} \sim a_R$.  The most massive disk we have created so far is $0.3M_E$,  from a collision with $M_T=7M_E$, $\gamma=0.1$, $\theta=50^{\circ}$, at $v_{imp}\sim v_{esc}$.  The disk material is iron-rich, and if it were assembled into a single satellite, its density would be 4.37 g/cm$^{3}$, comparable to the uncompressed density of the Earth.  This satellite is stable against tidal disruption, with $r_{sync} \sim 2 a_R$.  

We have, by necessity, calculated the upper limit on the mass of satellite that can be created in an impact.  Simulations of the accretion of impact-generated debris show that the disk need not form a single satellite, and that the mass of satellite(s) created additionally depends on the disk angular momentum and behavior of the disk inside the Roche zone \citep{IdaCanupStewart, SalmonCanup, SalmonCanup2014, Hyodo}.  A very promising avenue of future work would be to adapt the disk models and $N$-body accretion models to characterize moon formation in impact-generated disks of the type produced in our study.  This would also include looking at variations in disk composition (e.g., more metal-rich disks).  The disks produced in our study display a wide range of temperature and pressure conditions, with impacts between larger-mass planets yielding hotter disks.  Thus, the relationship between disk mass and final moon mass may change with $M_T$.  In future studies, we will report on disk temperature, pressure, and phase, so as to facilitate the construction of superearth disk models.

Our study of the types of exotic moon-forming impacts possible within exoplanetary systems has yielded promising initial results, relevant to the current efforts to observe exomoons. The models suggest that detectable rocky exomoons can be produced for a variety of impact conditions and may be associated with host planets of various sizes. Previously defined scaling relationships for expected moon masses produced in planetary collisions have been confined to the narrow range of cases specific to the Moon-forming impact at Earth. Hence, these terrestrial-focused scaling laws will need to be updated and expanded to accommodate the full range of natural collisions expected to occur in other planetary systems. As the available suite of simulation results grows, these laws will be defined more rigorously.

\section*{Acknowledgements}
Author Barr thanks the organizers of the 2014 Habitable Worlds Through Space and Time Conference, particularly John Debes, for soliciting an invited talk which motivated this work.  Part of this work was performed under the auspices of the U.S. Department of Energy by Lawrence Livermore National Laboratory under Contract DE-AC52-07NA27344. LLNL-JRNL-710518.




\bibliographystyle{mnras}
\bibliography{refs} 



\clearpage
\appendix
\section{Raw Data}

\begin{table}
	\centering
	\caption{Simulations Inputs and Outputs}
	\label{tab:sims}
	\begin{tabular}{lcccccccc} 
		\hline
		Name & $M_T/M_E$ & $\gamma$ & $v_{imp}/v_{esc}$ & $\theta$ &$M_{interact}/M_E$ & $M_{pl}/M_E$ & $M_o/M_E$ & $L_o/L_{EM}$ \\
		\hline
big\_earth	&	2.311	&	0.325	&	0.929	&	62.13	&	0.0151	&	2.152	&	0.127	&	3.202\\
pc\_earth	&	1.013	&	0.327	&	0.983	&	61.92	&	0.0068	&	0.939	&	0.061	&	0.854\\
rocky\_exo2&	0.266	&	0.146	&	0.986	&	50.55	&	0.0048	&	0.258	&	0.006	&	0.038\\
rocky\_exo3&	0.466	&	0.142	&	0.962	&	50.55	&	0.0083	&	0.458	&	0.006	&	0.065\\
rocky\_exo4&	2.324	&	0.123	&	0.862	&	50.55	&	0.0367	&	2.281	&	0.040	&	0.868\\
rocky\_exo5&	3.805	&	0.119	&	0.825	&	50.55	&	0.0582	&	3.768	&	0.032	&	0.854\\
rocky\_exo6&	7.208	&	0.115	&	0.774	&	50.55	&	0.1020	&	7.168	&	0.033	&	1.216\\
rocky\_exo7&	11.151&	0.111	&	0.736	&	50.55	&	0.1541	&	11.082&	0.058	&	2.791\\
rocky\_exo8&	18.066&	0.106	&	0.691	&	50.55	&	0.2387	&	17.987&	0.063	&	3.932\\
ser119	&	1.015	&	0.130	&	0.922	&	50.55	&	0.0170	&	0.998	&	0.014	&	0.215\\
velocity3	&	7.208	&	0.115	&	0.864	&	50.55	&	0.1020	&	7.068	&	0.112	&	5.182\\
velocity4	&	7.208	&	0.115	&	0.953	&	50.55	&	0.1020	&	7.048	&	0.071	&	3.216\\
velocity5	&	7.208	&	0.115	&	1.042	&	50.55	&	0.1020	&	6.802	&	0.311	&	6.997\\
velocity6	&	7.208	&	0.115	&	1.216	&	50.55	&	0.1020	&	6.571	&	0.003	&	0.107\\
velocity7	&	7.208	&	0.115	&	1.390	&	50.55	&	0.1020	&	6.489	&	0.005	&	0.192\\
velocity8	&	7.208	&	0.115	&	1.647	&	50.55	&	0.1020	&	6.438	&	0.002	&	0.089\\
angle1	&	7.208	&	0.115	&	0.774	&	20.70	&	0.6845	&	7.172	&	0.001	&	0.030\\
angle2	&	7.208	&	0.115	&	0.774	&	28.11	&	0.5710	&	7.176	&	0.000	&	0.012\\
angle3	&	7.208	&	0.115	&	0.774	&	36.09	&	0.3800	&	7.177	&	0.001	&	0.043\\
angle4	&	7.208	&	0.115	&	0.774	&	62.07	&	0.0293	&	7.102	&	0.090	&	4.478\\
angle5	&	7.208	&	0.115	&	0.774	&	70.46	&	0.0080	&	7.050	&	0.151	&	8.491\\
gamma1	&	7.126	&	0.196	&	0.778	&	50.55	&	0.1292	&	7.010	&	0.112	&	5.192\\
gamma2	&	7.131	&	0.321	&	0.778	&	50.55	&	0.1619	&	6.953	&	0.120	&	5.618\\
gamma3	&	7.264	&	0.451	&	0.771	&	50.55	&	0.1797	&	6.957	&	0.298	&	12.683\\
gamma4	&	7.518	&	0.442	&	0.757	&	50.55	&	0.1974	&	7.369	&	0.136	&	6.982\\
gamma5	&	7.176	&	0.137	&	0.775	&	50.55	&	0.1150	&	7.108	&	0.063	&	2.543\\
gamma6	&	7.112	&	0.258	&	0.779	&	50.55	&	0.1519	&	6.969	&	0.135	&	6.334\\
gamma7	&	7.182	&	0.386	&	0.775	&	50.55	&	0.1824	&	7.022	&	0.148	&	7.137	\\
		\hline
	\end{tabular}
	\end{table}	
		



\bsp	
\label{lastpage}
\end{document}